\documentclass[published]{JHEP3}

\usepackage{bbm}

\newcommand{\Tr}{\mathop{\rm Tr}\nolimits}
\newcommand{\SU}{\mathop{\rm SU}\nolimits}

\newcommand{\U}{\mathop{\rm {}U}\nolimits}

\def\l{\lambda}
\def\n{\nu}
\def\o{\omega}
\newcommand{\va}{|0,l\rangle}
\newcommand{\vc}{|0\rangle_{l+1} }
\newcommand{\vac}{|0\rangle_{\n} }

\title{Finite Chern-Simons matrix model --- algebraic approach}

\author{Larisa Jonke and Stjepan Meljanac\\ Theoretical Physics Division,
Rudjer Bo\v skovi\'c Institute\\ P.O. Box 180,
HR-10002 Zagreb, Croatia\\ E-mail: \email{larisa@thphys.irb.hr},
\email{meljanac@thphys.irb.hr}}

\abstract{We analyze the algebra of observables and the physical Fock
space of the finite Chern-Simons matrix model. We observe that the
minimal algebra of observables acting on that Fock space is
identical to that of the Calogero model. Our main result is  the
identification of the states in the l-th tower of the Chern-Simons
matrix model Fock space and the states of the Calogero model with
the interaction parameter $\n=l+1$. We describe quasiparticle and
quasihole states in the both models in terms of Schur functions,
and discuss some non-trivial consequences of our algebraic
approach.}


\received{November 07, 2001}

\accepted{January 09, 2002}

\JHEP{01(2002)008}

\begin{document}

\section{Introduction}

\looseness=-1
It is well known that the canonical  quantization of the system of
particles in a strong magnetic field  gives a natural realization
of non-commutative space. One can speculate whether it is
possible to describe a  real physical system --- Quantum Hall fluid
--- using quantum field theory on non-commutative plane. It was
conjectured~\cite{suss} that the Laughlin state of electrons at a
filling fraction\footnote{Actually, it was shown by Polychronakos
that owing to the quantum effects the corresponding filling
fraction is $1/(k+1)$.} $1/k$ was described by the
non-commutative version of the $\U(1)$ Chern-Simons theory at
level $k$. The fields in that theory were infinite matrices
corresponding to an infinite number of electrons on infinite
plane. Later, Polychronakos~\cite{pol1} proposed a regularized
version of the same model that could describe a finite number of
electrons localized on a plane. The complete minimal basis of
exact wavefunctions for the theory at an arbitrary level $k$ and
rank $N$ was given in ref.~\cite{hell}, and the coherent-state
representation was analyzed in ref.~\cite{s2}. Using the
properties of the energy eigenvalues of the Calogero
model~\cite{cal}, an orthogonal basis for the Chern-Simons (CS)
matrix model was identified~\cite{ks}.

The relation between the Calogero model and Quantum Hall (QH)
physics was investigated  using the algebraic
approach~\cite{bc,brink} and the collective-field
theory~\cite{iso}. In ref.~\cite{bc} it was conjectured  and then
proved in ref.~\cite{brink} that one could map   anyons  in the
lowest \mbox{Landau} level into the Calogero model, using the
complex representation of the $S_N$-extended Heisenberg algebra
underlying the Calogero model. On the other hand, it was shown in
ref.~\cite{iso} that the correlation functions of the QH edge
state and the Calogero model were related for the integer
interaction parameter $\n$. Also, the relation between the
Calogero model and the matrix model was established~\cite{p3}.
Finally, an interesting link between non-commutative CS theory
and QH fluid was provided using branes in massive type-IIA string
theory~\cite{d2brane}. Taking into consideration all these
relations between the Calogero model, the matrix model and QH
physics, one hopes that this intricate network of connections
between the apparently different physical systems will provide
useful insight into the underlying structure.

In this letter we analyze in detail the physical Fock space of
the finite  CS matrix model. We observe that the minimal algebra
of observables acting on that Fock space is identical to that of
the Calogero model, but we stress that the complete algebraic
structures are  different. Our main result is  the identification
of the states in the l-th tower of the CS matrix model Fock space
and the states of the Calogero model with the interaction
parameter $\n=l+1$. We discuss some mathematical and physical
consequences of this identification. Specialy, we describe
quasiparticle and quasihole excitations in the CS matrix model,
and using the identification, the corresponding excitations in
the Calogero model. We also discuss a possible extension of the
physical Fock space to include particles with the fractional
statistics. To make this paper self-contained, we add an appendix
with relevant expressions and results in the Calogero
model~\cite{clplb,b2,isakov}.

\section{CS matrix model}

\subsection{Introduction --- the physical Fock space}

Let us start from the action proposed in ref.~\cite{pol1}:
\begin{equation}\label{1}
S=\int dt\frac{B}{2}\Tr \left\{\varepsilon_{ab}\left(\dot X_a+
i\left[A_0,X_a\right]\right)X_b+2\theta A_0-\omega X_a^2\right\}+
\Psi^{\dagger}\left(i\dot\Psi -A_0\Psi\right).
\end{equation}
Here, $A_0$ and $X_a,\;a=1,2$, are $N\times N$ hermitean matrices
and $\Psi$ is a complex $N$-vector. The eigenvalues of the
matrices $X_a$ represent the coordinates of electrons, $A_0$ is a
gauge field, and $\Psi$ acts like a boundary field. We  choose
the gauge $A_0=0$ and impose the equation of motion for $A_0$  as
a constraint:
\begin{equation}\label{2}
-iB\left[X_1,X_2\right]+\Psi\Psi^{\dagger}=B\theta\,.
\end{equation}
The trace part of eq.~(\ref{2}) gives
\begin{equation}\label{3}
\Psi^{\dagger}\Psi=NB\theta\,.
\end{equation}
Notice that the commutators have so far been classical matrix
commutators. After quantization, the matrix elements of $X_a$ and
the components of $\Psi$ become operators satisfying the
following commutation relations:
\begin{eqnarray}\label{4}
\left[\Psi_i,\Psi_j^{\dagger}\right]&=&\delta_{ij}\,,\nonumber \\
\left[\left(X_1\right)_{ij},\left(X_2\right)_{kl}\right]&=&\frac{i}{B}
\delta_{il}\delta_{jk}\,.
\end{eqnarray}
It is convenient to introduce the operator $A=\sqrt{B/2}(X_1+iX_2)$
and its hermitean conjugate $A^{\dagger}$ obeying the following
commutation relations:
\begin{equation}\label{5}
\left[A_{ij},A^{\dagger}_{kl}\right]= \delta_{il}\delta_{jk}\,,\qquad
\left[A_{ij},A_{kl}\right]= \left[A^{\dagger}_{ij},A^{\dagger}_{kl}
\right]=0\,.
\end{equation}
Then, one can write the hamiltonian of the model at hand as
\begin{equation}\label{6}
H=\omega\left(\frac{N^2}{2}+\Tr (A^{\dagger}A)\right)=
\omega\left(\frac{N^2}{2}+\mathcal{N}_A\right),
\end{equation}
$\mathcal{N}_A$ being the total number operator associated with
$A$'s. Upon quantization, the constraints~(\ref{2}) become the
generators of unitary transformations of both $X_a$ and $\Psi$.
The trace part~(\ref{3}) demands that (the l.h.s.\ being the
number operator for $\Psi$'s) $B\theta\equiv l$ be quantized to
an integer. The traceless part of the constraint~(\ref{2})
demands that the wavefunction be invariant under $\SU(N)$
transformations, under which $A$ transforms in the
adjoint\footnote{Note that as $A$ transforms in the reducible
representation $(N^2-1)+1$, with the singlet $B_1\equiv \Tr A$,
one can introduce a pure adjoint  representation as $\bar
A_{ij}=A_{ij}-\delta_{ij}B_1/N$. This slightly modifies the
commutator~(\ref{5}), and completely decouples $B_1$ from the
Fock space. Physically, this correspond to the separation of the
centre-of-mass coordinate as it has been done for the Calogero
model in ref.~\cite{clplb}. For the sake of simplicity, this will
not be done here.} and $\Psi$ in the fundamental representation.

Energy eigenstates will be $\SU(N)$ singlets; generally, some
linear combinations of terms with at least $lN(N-1)/2$
$A^{\dagger}$ operators and $Nl$ $\Psi^{\dagger}$ fields.
Explicit expressions for the wavefunctions were written in
ref.~\cite{hell}:
\begin{equation}\label{8}
|\Phi\rangle=\prod_{j=1}^{N}(\Tr A^{\dagger j})^{c_j}
C^{\dagger l}|0\rangle\,,
\end{equation}
where
\begin{equation}
C^{\dagger}\equiv\varepsilon^{i_1\cdots i_N}\Psi^{\dagger}_{i_1}
(\Psi^{\dagger}A^{\dagger})_{i_2}\cdots (\Psi^{\dagger}
A^{\dagger N-1})_{i_N}\,,
\end{equation}
and  $A_{ij}|0\rangle=\Psi_i|0\rangle=0$.

The system contains $N^2+N$ oscillators coupled by $N^2-1$
constraint equations in the traceless part of eq.~(\ref{2}).
Effectively, we can describe the system wih $N+1$ independent
oscillators. Therefore, the physical Fock space that consists  of
all $\SU(N)$-invariant states can be spanned by $N+1$
algebraically independent operators: $B_n^{\dagger}\equiv \Tr
A^{\dagger n}$ with $n=1,2,\dots,N$, and $C^{\dagger}$. The
operators $B_k^{\dagger}$ for $k>N$ can be expressed as a
homogeneous polynomial of total order k in $\{B_1^{\dagger},\dots,
B_N^{\dagger}\}$, with constant coefficients which are common to
all operators $A^{\dagger}$~\cite{macf}. Since
\begin{equation}\label{vacuum}
\Tr A^kC^{\dagger l}|0\rangle\equiv B_kC^{\dagger l}|0\rangle=0\,,\qquad
\forall k\,,\ \forall l\,,
\end{equation}
the state $C^{\dagger l}|0\rangle\equiv\va$ can be interpreted as a
ground state ---  vacuum with respect to all operators $B_k$. Note
that the vacuum is not normalized to one, i.e.\ $\langle 0,l\va\neq 1$.
The whole physical Fock space can  be decomposed into  towers
(modules) built on the ground states with different $l$:
$$
F_{\mathrm{phys}}^{\mathrm{CS}}=\sum_{l=0}^{\infty}
F_{\mathrm{phys}}^{\mathrm{CS}} (l)= \sum_{l=0}^{\infty}\left\{\prod
B_k^{\dagger n_k}\va\right\}.
$$
Clearly, the states from different towers are mutually orthogonal.

The action of the hamiltonian~(\ref{6}) on the states in the
physical Fock space is
\begin{equation}\label{act}
H\prod B_k^{\dagger n_k}\va=\o\left[\frac{N^2}{2}+ \sum_kkn_k+
l\pmatrix{N\cr 2}\right]\prod B_k^{\dagger n_k}\va\,,
\end{equation}
and the ground-state energy is $E_0(l)=\o[N^2+lN(N-1)]/2$.
Comparing this result with the known one for the Calogero model
(see eq.~(\ref{H}) in appendix~\ref{appA}) we see that the
spectra of the two models are identical provided $\n=l+1$. This
has already been noticed in refs.~\cite{ks,p3}. However, from the
identification of spectra it only follows
$$
\left(\prod B_i^{\dagger n_i}\right)^{\mathcal{N}}\va=
\sum\left(\prod B_k^{\dagger n_k}\right)^{\mathcal{N}}\vc\,,
$$
where the r.h.s.\ of this relation is, in general,  a sum of
different terms of total order $\mathcal{N}$  in the observables
of the Calogero model. We use the same letter to denote
observables in  both models. In CS matrix model, $B_n=\Tr A^n$
and in the Calogero model, $B_n=\sum_ia_i^n$, but from the
context it should be clear what $B_n$ represents. The vacuum  in
the CS matrix model is denoted as $\va$ and the vacuum in the
Calogero model is denoted as $\vac$ (see appendix~\ref{appA}).

\subsection{Algebraic structure of the CS matrix model}

Using $[A_{ij},B_n^{\dagger}]=n(A^{\dagger n-1})_{ij}$, we find
a general expression for the commutators between observables:
\begin{equation}\label{B}
[B_m,B_n^{\dagger}]=n\sum_{r=0}^{m-1}\Tr (A^rA^{\dagger n-1}A^{m-r-1})=
m\sum_{s=0}^{n-1}\Tr (A^{\dagger s}A^{m-1}A^{\dagger n-s-1})\,.
\end{equation}
One can normally order the r.h.s.\ of eq.~(\ref{B}) using the
recurrent relation
\begin{equation}\label{no}
\Tr (A^rA^{\dagger n-1}A^{m-1})=\Tr(A^{r-1}A^{\dagger n-1}A^m)+
\sum_{s=0}^{n-2}\Tr(A^{r-1}A^{\dagger s})\Tr(A^{\dagger n-s-2}
A^{m-r-1})\,.
\end{equation}
With the formal mapping $\Tr(A^rA^{\dagger s}A^k)\rightarrow
\sum_ia_i^ra_i^{\dagger s}a_i^k$, the relation~(\ref{B}) is
identical to eq.~(\ref{cB}) for the Calogero model. Also, the
recurrent relation~(\ref{no}) has its  counterpart in the
Calogero model with $\n=1$, with the same formal mapping. In
order to close the algebra~(\ref{B}), one should include
observables of the type $B^{\alpha}_{m,n}=\Tr(A^{\dagger
i_1}A^{i_2}A^{\dagger i_3}\cdots)$, where $m$ is the number of
$A^{\dagger}$'s  and $n$ is the number of $A$'s, in the trace.
This algebra $\mathcal{B}_N^{\mathrm{CS}}$ is a polynomial
generalization of the Lie algebra. We stress that this algebra is
larger than the corresponding $\mathcal{B}_N^{\mathrm{Cal}}$
algebra in the Calogero model described in detail in
ref.~\cite{clplb}, since in the CS matrix model we have more than
one invariant of order $(m,n)$, owing to matrix multiplication.
In other words, in the Calogero model there are more algebraic
relations connecting the trace invariants for fixed $N$.

In this paper we restrict ourselves  to the minimal algebra
including only  observables of the type $B_n$ and $B_n^{\dagger}$,
defined with the following relations (including corresponding
hermitean conjugate relations):
\begin{equation}\label{D}
[B_{i_1},[B_{i_2},[\dots,[B_{i_n},B_n^{\dagger}]\dots]]=n!
\prod_{\alpha=1}^n i_{\alpha}B_{I-n}\,,
\end{equation}
where $I=\sum_{\alpha=1}^ni_{\alpha}$ and $i_1,\dots,i_n,n=1,2,\dots, N$.
These relations can be viewed as a generalization of triple operator
algebras defining para-Bose and para-Fermi statistics~\cite{trip}.
The identical successive commutators
relations~(\ref{D}) holds for the observables acting on the
$S_N$-symmetric Fock space of the Calogero model~\cite{clplb}.
The action of $B_k$ on any state in the Fock space is of the form:
\begin{equation}\label{gV}
B_k\prod_i B_i^{\dagger n_i}\va=\sum\left(\prod B_j^{\dagger n_j}\right)
^{\mathcal{N}-k} \va\,,\qquad \mathcal{N}=\sum_i in_i\geq k\,,
\end{equation}
In order to calculate the precise form of the r.h.s.\ of
eq.~(\ref{gV}) we apply the hermitean conjugate relation
eq.~(\ref{D}) on the l.h.s.\ of eq.~(\ref{gV}) shifting the
operator $B_k$ to the right, at least for one place. We repeat
this iteratively as long as the number of $B^{\dagger}$'s on the
right from $B_k$ is larger or equal to the index $k$. For $k>\sum
n_i$, one directly calculates a finite set of relations
from~(\ref{5}), the so-called generalized vacuum conditions. We
show that  the \emph{minimal} set of generalized vacuum conditions
needed to completely define the representation of the
algebra~(\ref{D}) on the Fock space is
\begin{eqnarray}\label{V}
B_2B_2^{\dagger}\va &=&2N(N+lN-l)\va\,,\nonumber \\
B_3B_3^{\dagger}\va &=&3N[N^2+1+l(N-1)(2N-1)+l^2(N-1)(N-2)]\va=y\va\,,
\nonumber \\
B_3B_3^{\dagger 2}\va &=& 54\{(l+1)B_1^{\dagger}B_2^{\dagger}+[N+(N-2)l+
y/27]B_3^{\dagger}\} \va\,.
\end{eqnarray}
Namely, the operators $B_2,\;B_3$ and hermitean conjugates play a
distinguished role in the algebra, since all other operators
$B_n,\;B_n^{\dagger}$ for $n\geq 4$ can be expressed as
successive commutators~(\ref{D}), using only $B_2,\;B_3$ and
their hermitean conjugates. Therefore, one can  derive all other
generalized vacuum conditions using~(\ref{D}) and~(\ref{V}).

The relations~(\ref{D}) and the generalized vacuum
conditions~(\ref{V}) represent the minimal algebraic structure
defining the complete physical Fock space representation. Using
these relations one can calculate the action of $B_n$ on any
state in the physical Fock space, and calculate all matrix
elements (scalar product) of the form $\langle 0,l|(\prod
B_{i}^{n_{i}}) (\prod B_{j}^{\dagger n_{j}})|0,l\rangle$, up to
the norm of the vacuum.

\subsection{Relation to the Calogero model}

Our main observation is that the generalized vacuum conditions
for the CS matrix model (\ref{V}) and for the Calogero
model~(\ref{cV}) coincide for $\n=l+1$. This follows entirely
from the common algebraic structure~(\ref{D}) and the structure
of the vacuum conditions. As we have already said, the algebraic
relations~(\ref{D}) and the generalized vacuum conditions
uniquely determine the action of observables on the Fock space, so
we conclude that
\begin{equation}\label{id1}
B_k\left(\prod B_i^{\dagger n_i}\right)\va =B_k\left(\prod B_i^{\dagger n_i}
\right)\vc\,,
\end{equation}
and that all scalar products in $F_{\mathrm{phys}}^{\mathrm{CS}}$ can be
identified with the corresponding matrix elements in the Fock
space of the Calogero model:
\begin{equation}\label{id2}
\langle 0,l|\left(\prod B_{i}^{n_{i}}\right)\left(\prod
B_{j}^{\dagger n_{j}} \right)|0,l\rangle= \;_{l+1}\langle 0|
\left(\prod B_{i}^{n_{i}}\right)\left(\prod B_{j}^{\dagger n_{j}}\right)
\vc\,.
\end{equation}
Note that the vacuum is unique in both models, so we set $\va
=\vc$, up to the norm. Now we can identify all states in
$F_{\mathrm{phys}}^{\mathrm{CS}}(l)$ with the corresponding
states in $F_{\mathrm{symm}}^{\mathrm{Cal}}(l+1)$ as
\begin{equation}\label{id3}
\left(\prod B_k^{\dagger n_k}\right)\va =\left(\prod B_k^{\dagger n_k}
\right)\vc\,.
\end{equation}
One can prove these results in different ways. For example, we
can restrict ourselves to the subspace of the Fock space
generated by $\{B_1^{\dagger},B_2^{\dagger}, B_3^{\dagger}\}$ and
prove the relations~(\ref{id1}),~(\ref{id2}) and~(\ref{id3}) in
this subspace, by straightforward calculation and by induction.
Then, using the algebraic relation~(\ref{D}) we simply generate
the rest of the needed results on the full (physical) Fock space.
The other way to obtain our result is to notice that from the
equivalence of spectra it follows $\n=l+1$. Also, from the fact
that the algebraic relations~(\ref{D}) hold for both models we
know that the states in the Calogero model and the CS matrix
model are identical, but we have no information on the relation
between $l$ and $\n$. The combination of these two results leads
again  to the identification of the states in
$F_{\mathrm{phys}}^{\mathrm{CS}}(l)$ and
$F_{\mathrm{symm}}^{\mathrm{Cal}}(l+1)$, eq.~(\ref{id3}). We
point out that our algebraic proof of
eqs.~(\ref{id1}),~(\ref{id2}) and~(\ref{id3}) relies only on the
identical algebra~(\ref{D}) and the identical minimal set of
generalized vacuum conditions for both models, and not on the
structure of their hamiltonians.

This identification of the states in the CS matrix model with the
corresponding states in the Calogero model is non trivial, from
both mathematical and physical point of view. An important
consequence of our analysis is that the operators $a_i$, defined
in eq.~(\ref{defa}), can be interpreted as elements belonging to
the spectrum of the matrix $A$. Namely, they are solutions of the
polynomial equation $\det |A-a \cdot \mathbbm{1}_{N\times N}|=0$,
i.e.\ $\sum_k\alpha_ka^k=0$, where $\alpha_k$ are
operators commuting among themselves and with $a$'s.
Let us assume that there are $N$ different, mutually commuting solutions
$a_1,\dots,a_N$; then $\Tr A^n=\sum a_i^n$. This pure algebraic
relation should be applied to the ground states, according to our
results:
$$
\Tr A^n\va=\sum_i a_i^n\vc=0\,,\qquad \mbox{and}\qquad
\prod(\Tr A^{\dagger k})^{n_k} \va =\prod\left(\sum_i a_i^{\dagger k}
\right)^{n_k}\vc\,.
$$
Therefore, we conclude that, for consistency reasons, the
operators $a_i, a_j^{\dagger}$ have to satisfy
$[a_i,a_j^{\dagger}]=-(l+1)K_{ij},\;i\neq j$.

Also, as a consequence of this identification, we generate an
infinite number of non-trivial identities in the following way.
Using the relations valid for $m\leq n$
\begin{eqnarray}
\left[B_m,B_n^{\dagger}\right]^{\mathrm{CS}} &=& \sum_{l=1}^mc_{lmn}(1)
\Tr(A^{\dagger n-l} A^{m-l})\,,\nonumber \\
\left[B_m,B_n^{\dagger}\right]^{\mathrm{Cal}} &=& \sum_{l=1}^mc_{lmn}(\n)
\sum_{i=1}^N a_i^{\dagger n-l}a_i^{m-l}\,,
\end{eqnarray}
where $c_{lmn}(\n)$ are coefficients depending on all indices and
$\n$, we find an infinite number of identities:
\begin{eqnarray}
[B_m,B_n^{\dagger}]^{\mathrm{CS}}\va &=& [B_m,B_n^{\dagger}]^{\mathrm{Cal}}
\vc\,,\nonumber \\
\sum_{k=1}^mc_{kmn}(1)\Tr(A^{\dagger n-k}A^{m-k})\va &=& c_{mmn}(l+1)
B^{\dagger}_{n-m}\vc\,.
\end{eqnarray}
For example, from $[B_2,B_{n+2}^{\dagger}]$, $n>0$:
\begin{equation}
\sum_{k=0}^{N-2}(N-k-1)(-)^kS_{n-k,1^k}=(N-1)m_{(n)}+\sum_{r=1}^{[n/2]}
m_{(n-r,r)}\,,\nonumber
\end{equation}
where $S_{\l}$ is Schur function and $m_{\l}$ is symmetric monomial
function, see ref.~\cite{mac}.

\section{Quasiparticles and quasiholes}

The low-lying excitations in the CS matrix model can be described
in terms of quasiparticles and quasiholes~\cite{pol1}. We can
identify these states in a way analogous  to that for particles
and holes of a Fermi sea. One quasiparticle obtained by exciting
a ``particle'' at Fermi level by energy amount $\Delta E=n \o$ is
\begin{equation}\label{qp}
\pi^{\dagger}(n)\va=C^{\dagger (l-1)}\times \varepsilon^{i_1\cdots i_N}
\Psi^{\dagger}_{i_1}\cdots (\Psi^{\dagger}A^{\dagger N-2})_{i_{N-1}}
(\Psi^{\dagger}A^{\dagger N-1+n})_{i_N}|0\rangle\,.
\end{equation}
One quasihole excitation is obtained by creating a gap inside the
QH droplet with energy increase $\Delta E=(N-k)\o$
\begin{equation}\label{qh}
\chi^{\dagger}(k)\va=C^{\dagger (l-1)}\times \varepsilon^{i_1\cdots i_N}
\Psi^{\dagger}_{i_1}\cdots (\Psi^{\dagger}A^{\dagger k-1})_{i_{k}}
(\Psi^{\dagger}A^{\dagger k+1})_{i_{k+1}} \cdots(\Psi^{\dagger}
A^{\dagger N})_{i_N}|0\rangle\,.
\end{equation}
Observe that $\chi^{\dagger}(N-1)=\pi^{\dagger}(1)$. To see that
the charge, i.e.\ the  particle number of a quasihole is
quantized as $1/l$, we simply observe that removing one particle
(removing $(\Psi^{\dagger}A^{\dagger k})^l$) from the ground state
is equivalent to creation of $l$ quasiholes.
\begin{eqnarray*}
P^{\dagger}(N)P(k)\va &=& \chi^{\dagger l}(k)\va
\nonumber\\
&=& \left[C^{\dagger(l-1)}\times\varepsilon^{i_1\cdots i_N}\Psi^{\dagger}_{i_1} \cdots
(\Psi^{\dagger}A^{\dagger k-1})_{i_{k}}(\Psi^{\dagger}
A^{\dagger k+1})_{i_{k+1}} \cdots(\Psi^{\dagger}A^{\dagger N})_{i_N}
\right]^l|0\rangle\,.
\end{eqnarray*}

We wish to write the expressions~(\ref{qp}) and~(\ref{qh}) in our
basis of the physical Fock space, so we introduce the following
functions:
\begin{eqnarray*}
{\sigma}_n(B_1^{\dagger},\dots,B_N^{\dagger}) &=&
\frac{\varepsilon^{i_1\cdots i_N}\Psi^{\dagger}_{i_1}\cdots
(\Psi^{\dagger}A^{\dagger N-2})_{i_{N-1}}
(\Psi^{\dagger}A^{\dagger N-1+n})_{i_N}}{C^{\dagger}}\,,\\
{\sigma}_{1^{N-k}}(B_1^{\dagger},\dots,B_N^{\dagger}) &=&
\frac{\varepsilon^{i_1\cdots i_N}\Psi^{\dagger}_{i_1}\cdots
(\Psi^{\dagger}A^{\dagger k-1})_{i_{k}}(\Psi^{\dagger}
A^{\dagger k+1})_{i_{k+1}} \cdots(\Psi^{\dagger}A^{\dagger N})_{i_N}}
{C^{\dagger}}\,.
\end{eqnarray*}
Now, quasiparticle and quasihole excitations are written as
$\pi^{\dagger}(n)\va={\sigma}_n(B_1^{\dagger},\dots,B_N^{\dagger})$
$\va$ and $\chi^{\dagger}(k)\va ={\sigma}_{1^{N-k}}(B_1^{\dagger},
\dots,B_N^{\dagger})\va$, respectively. In order to precisely determine
the functions ${\sigma}_{\lambda}$, we use the mapping from the
CS matrix model to the Calogero model and the fact that the
operators $a_i^{\dagger}$, defined in eq.~(\ref{defa}), belong to
the spectrum of the matrix $A^{\dagger}$. From the definition of
Schur functions $S_{\lambda}$, and using the notation of
ref.~\cite{mac}, we have
\begin{eqnarray}
{\sigma}_n(B_1^{\dagger},\dots,B_N^{\dagger})\va&=&
S_n(a_1^{\dagger},\dots,a_N^{\dagger})\vc=
\sum_{|\lambda|=n}z_{\lambda}^{-1}B^{\dagger}_{\l_1}\cdots
B^{\dagger}_{\l_N} \vc\,, \label{sch}\\
{\sigma}_{1^{N-k}}(B_1^{\dagger},\dots,B_N^{\dagger})\va&=& S_{1^{N-k}}
(a_1^{\dagger},\dots,a_N^{\dagger})\vc = \sum_{|\lambda|=N-k}
\varepsilon_{\lambda} z_{\lambda}^{-1}B^{\dagger}_{\l_1}\cdots
B^{\dagger}_{\l_N}\vc\,.\nonumber
\end{eqnarray}
Here the summation goes over all partitions $\l=(\l_1,\dots,\l_N),\;
\l_1\geq\l_2 \geq\cdots\geq\l_N\geq 0$ of given weight $|\lambda|=
\sum_i\l_i$ and lenght $l(\l)$. The coefficients in relations~(\ref{sch})
are $z_{\lambda}= \sum_ii^{m_i}m_i!$, where $m_i$ is the number of $i$'s in
the partition $\l$, and $\varepsilon_{\lambda}=(-)^{|\l|-l(\l)}$.
The same expression for the quasihole wave function in the
Calogero model has been written down in ref.~\cite{plmpl} and,
also, the norm of this state has been given:
$$
_{\n}\langle 0|\chi(k)\chi^{\dagger}(k)\vac =\pmatrix{N\cr k}
\prod_{i=0}^{k-1}[1+\n i]\,.
$$
The partition $\l=(\l_1,\dots,\l_N)$ can be represented by the
Young tableau with $N$ rows, each having $\l_i$ boxes. In the
Calogero model, the quasiparticle $\pi^{\dagger}(n)$ is then
represented by one row with $n$ boxes and quasihole
$\chi^{\dagger}(n)$ by a single column with $N-n$ boxes.

In this representation it is evident that there is no fundamental
distinction between quasiparticles and quasiholes for finite $N$,
since they obviously represent the same type of excitations in
two different bases. Using the properties of Schur function we
can write a simple relation connecting these two bases:
$$
S_n=\sum_{l=1}^n(-)^{l+n}\sum_{i_1,\dots,i_l=1}^NS_{1^{i_1}}\cdots
S_{1^{i_l}}= \sum_{l=1}^n(-)^{l+n}\sum_{|\mu|=l}
\pmatrix{\mu_1+\cdots+\mu_N \cr \mu_1, \dots,\mu_N}S_1^{\mu_1}\cdots
S_{1^N}^{\mu_N}\,,
$$
and vice versa
$$
S_{1^n}=\sum_{l=1}^n(-)^{l+n}\sum_{i_1,\dots,i_l=1}^NS_{i_1}\cdots
S_{i_l}= \sum_{l=1}^n(-)^{l+n}\sum_{|\mu|=l}
\pmatrix{\mu_1+\cdots+\mu_N \cr \mu_1, \dots,\mu_N}S_1^{\mu_1}\cdots
S_N^{\mu_N}\,,
$$
where  summations over $i_{\alpha}$ and $\mu$ are subject to condition
$\sum_{\alpha=1}^li_{\alpha}=\sum_{j=1}^Nj\mu_j=n$.
These relations hold for $\sigma_{\l}$ funtions also, i.e.\
we can apply these relations in both models.

\section{Outlook and discussion}

It is important to note that
the whole picture is consistent only if the matrix $A$ is not diagonal.
If we assume $A_{ij}=a_i\delta_{ij}$, then $[a_i,a_j^{\dagger}]
=\delta_{ij}$, and that corresponds to the $l=-1$ tower which does not
exist in the CS matrix model (although the
$\n=l+1=0$ case corresponds to $N$ bosons in
the Calogero model). There is also another way to see this inconsistency.
If we diagonalize the matrix $A$,
the ground states factorize as $\va\sim (\prod\Psi_i^{\dagger})^l\prod_{i<j}
(a_i^{\dagger}-a_j^{\dagger})^l$ and, generally, the vacuum conditions are
not satisfied any longer,
$(\sum_i a_i^n)\prod(a_i^{\dagger}-a_j^{\dagger})^l\vac
\neq 0$ for $n\leq l$.

Similar arguments can be applied to the states of the form
$(\sum_i a_i^{\dagger k})^{n_k}
\prod(a_i^{\dagger}-a_j^{\dagger})^l\vac$ with $[a_i,a_j^{\dagger}]
=\delta_{ij}$, eqs.~(11) and (12) in ref.~\cite{hell}, which correspond to
the  Laughlin states in a Fock space. For $l$ even, the above states are
completely symmetric and can be written as $(\sum_i a_i^{\dagger k})^{n_k}
\vac$, and for any odd $l$, the states are $(\sum_i a_i^{\dagger k})^{n_k}
\prod(a_i^{\dagger}-a_j^{\dagger})\vac$. Hence, all $l$-even states
reduce to a Bose tower with $l=0$, and all $l$-odd states reduce
to a Fermi tower with $l=1$. \pagebreak[3] Moreover, the vacuum
conditions are not satisfied. So, there is no one-to-one mapping
between the CS matrix model states and the above Fock-space
states. This leaves us with an unanswered question about the
relation between QH physics and the finite CS matrix model, as
was also observed in ref.~\cite{opet}.

Using the mapping from F$_{\mathrm{ph}}^{\mathrm{CS}}(l)$ to
$F^{\mathrm{Cal}}_{\mathrm{symm}}(l+1)$ we can show that there exists
a mapping from the free Bose oscillators
$\{b_1,b_2,\dots,b_N,c_0\}$ to the observables in the CS matrix
model $\{B_1,B_2,\dots,B_N,C\}$, and vice versa~\cite{b2}. This
mapping offers a natural orthogonal basis in the physical Fock
space $\{\prod(b_k^{\dagger})^{n_k}\va\}$, an alternative to the
orthogonal basis in terms of Jack polynomials proposed in
ref.~\cite{ks}

Also, the dynamical symmetry generators operating on degenerate
states in a fixed tower of states
F$_{\mathrm{ph}}^{\mathrm{CS}}(l)$ are the same as those found in
the Calogero model with $\n=l+1$~\cite{clplb,b2}. Moreover, one
can introduce additional generators acting between different
towers and so describe  the larger dynamical symmetry operating
on all degenerate states in the CS matrix model.

The relation between the Calogero model and the physics of anyons
in the lowest Landau level has been described in
ref.~\cite{brink}, and this relation is not restricted to the
integer values of the interaction parameter in the Calogero
model. It would be interesting to somehow enlarge this finite CS
matrix model to describe also particles with fractional
statistics. This can be easily  done in the Fock-space approach.

\looseness=-1
We can define a new $\SU(N)$ invariant operator
$D^{\dagger}=C^{\dagger 1/q}$ such that $\mathcal{N}_D=q
\mathcal{N}_C$,  $\mathcal{N}_D$ and $\mathcal{N}_C$ being the
number operators for $D$'s and $C$'s, respectively. The enlarged
Fock space $F_{\mathrm{phys}}^{\mathrm{CS,q}}$ is a space of all
states of the type $\prod_k(B_k^{\dagger})^{n_k}D^{\dagger l}
\vac$, for $l$ integer. From the action of the
hamiltonian~(\ref{6}) on that Fock space we obtain $\n=l/q+1$.
Also, the generalized vacuum conditions~(\ref{V}) hold for any
rational number $l\rightarrow l/q$, and therefore all
identifications eqs.~(\ref{id1}),~(\ref{id2}) and~(\ref{id3})
between the CS matrix and the Calogero model are true for any
rational $\n=l/q+1$. This construction is an allowed extension of
Fock space for rational $\n$, but we point out that the dynamical
origin of such a picture is still missing.

In ref.~\cite{suss} it was conjectured that the theory for QH
fluid with the filling fraction $n/k$ would be level $k$ $\U(n)$
non-commutative CS theory. In the string theory approach level
$k$ $\U(n)$ non-commutative CS theory corresponds to the
configuration of $n$ D2-branes and $2k$ D8-branes in a
background  $B$-field~\cite{d2brane}. Even in this approach, the
dynamical description of the hierarchy structure and the mass gap
of excitations in QH fluid is still missing.

In conclusion, we have  discussed the  Fock-space picture  in
detail and  elucidated the $A$-representation of the CS matrix
model. We have  precisely formulated the connection between the
states in the Fock spaces of the CS matrix and the Calogero model.
We stress that although the  models have similar Fock spaces and
a common minimal algebraic structure, the complete algebraic
structures are quite different. An important consequence of our
analysis is that the operators $a_i$, elements of $S_N$-extended
Heisenberg algebra, can be interpreted as elements belonging to
the spectrum of the matrix $A$.

\acknowledgments

We would like to thank D.~Svrtan for useful discussions. This
work was supported by the Ministry of Science and Technology of
the Republic of Croatia under contract No.~00980103.

\appendix

\section{Calogero model}\label{appA}

The hamiltonian of the (rational) Calogero model describes $N$
identical particles (bosons) interacting through an inverse
square interaction subjected to a common confining harmonic force:
\begin{equation}\label{c1}
H=-\frac{\hbar^2}{2m}\sum_{i=1}^N\frac{\partial^2}{\partial x_i^2}
+\frac{m\omega^2}{2}\sum_{i=1}^N x_i^2 +\frac{\n(\n-1)\hbar^2}{2m}
\sum_{i\neq j}^N\frac{1}{(x_i-x_j)^2}\,.
\end{equation}
After performing a similarity transformation on the
hamiltonian~(\ref{c1}), we obtain the reduced hamiltonian acting
on the space of symmetric functions ($m=\hbar=1$):
\begin{equation}\label{c7}
H'=\prod_{i<j}^N|x_i-x_j|^{\n}H\prod_{i<j}^N|x_i-x_j|^{-\n}=
\sum_{i=1}^Na_i^{\dagger}a_i+E_0\,,
\end{equation}
where the ground-state energy is $E_0=\o N[1+(N-1)\n]/2$. Here,
we have introduced the creation and annihilation operators
\begin{eqnarray} \label{defa}
a_i^{\dagger}&=&\frac{1}{\sqrt{2}}\left(-\partial_i-\n\sum_{j,j\neq i}^N
\frac{1}{x_i-x_j}(1-K_{ij})+\o x_i\right),\nonumber \\
a_i&=&\frac{1}{\sqrt{2}}\left(\partial_i+\n\sum_{j,j\neq i}^N
\frac{1}{x_i-x_j}(1-K_{ij})+\o x_i\right),
\end{eqnarray}
satisfying the following commutation relations
\begin{equation}\label{c6}
[a_i,a_j]=[a_i^{\dagger},a_j^{\dagger}]=0\,,\qquad [a_i,a_j^{\dagger}]=
\left(1+\n\sum_{k=1}^NK_{ik}\right)\delta_{ij}-\n K_{ij}\,.
\end{equation}
The elementary generators $K_{ij}$ of the symmetry group $S_N$
exchange labels $i$ and $j$. The Fock space representation is
defined by $ a_i|0\rangle=0$ and $K_{ij}|0\rangle=|0\rangle$. The
physical Fock space is defined by $S_N$-symmetric states
$F^{\mathrm{Cal}}_{\mathrm{symm}}=\{\prod_{n_k}B_k^{\dagger
n_k}|0\rangle\}$, where $B_k=\sum_ia_i^k$ are collective,
$S_N$-symmetric operators. Then one can write
\begin{equation}\label{H}
H'\prod_{n_k}B_k^{\dagger n_k}\vac=[E_0+\o \sum_{k=1}^Nkn_k]
\prod_{n_k}B_k^{\dagger n_k}\vac\,.
\end{equation}

The $S_N$-symmetric observables $B_n^{\dagger}$ creating the
symmetric Fock space satisfy the following commutation relations:
\begin{eqnarray}\label{cB}
[B_m,B_n] &=& [B_m^{\dagger}, B_n^{\dagger}]=0,\nonumber \\
{[B_m,B_n^{\dagger}]} &=& n\sum_{r=0}^{m-1}\sum_{i=1}^Na_i^r
a_i^{\dagger n-1} a_i^{m-r-1}= m\sum_{s=0}^{n-1}\sum_{i=1}^N
a_i^{\dagger s}a_i^{m-1}a_i^{n-s-1}\,.
\end{eqnarray}
Including the observables of the type $B_{m,n}=\sum a_i^{\dagger
m}a_i^n,\; m+n\leq N$ into consideration leads to the closed
polynomial Lie algebra $\mathcal{B}_N^{\mathrm{Cal}}$ described
in ref.~\cite{clplb}.

As we have already mentioned, the same, universal
relation~(\ref{D}) holds also in the Calogero model for the
observables $B_n$ acting on the $S_N$-symmetrical Fock space:
$$
[B_{i_1},[B_{i_2},[\dots,[B_{i_n},B_n^{\dagger}]\dots]]=
n!\prod_{\alpha=1}^n i_{\alpha}B_{I-n}\,,
$$
where $I=\sum_{\alpha=1}^ni_{\alpha}$ and $i_1,\dots,i_n,n=1,2,\dots, N$.
The representation of algebra~(\ref{D}) in
$F_{\mathrm{sym}}(\n)$ is completely characterized by the minimal
set of the generalized vacuum conditions:
\begin{eqnarray}\label{cV}
B_2B_2^{\dagger}\vac &=& 2N(N\n+1-\n)\vac\,,\nonumber \\
B_3B_3^{\dagger}\vac &=& 3N[2(\n-1)^2+\n-3N\n(\n-1)+N^2\n^2]\vac=y\vac\,,
\nonumber \\
B_3B_3^{\dagger 2}\vac &=& 54\{\n B_1^{\dagger}B_2^{\dagger}+[N+(N-2)(\n-1)
+y/27]B_3^{\dagger}\}\vac\,.
\end{eqnarray}
In the same way as in the CS matrix model one can show that all
other generalized vacuum conditions can be calculated using the
algebra~(\ref{D}) and relations~(\ref{cV}). The action on $B_n$
on any state in the $S_N$-symmetric Fock space, and all matrix
elements of the form $_{\n}\langle 0|(\prod B_{i}^{n_{i}})(\prod
B_{j}^{\dagger n_{j}})\vac$ are uniquely determined by~(\ref{D})
and~(\ref{cV}).

\end{document}